\nonstopmode
\documentclass[a4paper]{jpconf}
\usepackage{graphicx}
\usepackage{amsmath}
\usepackage{graphicx}

\newcommand{\bra}[1]{\langle #1|}
\newcommand{\ket}[1]{|#1\rangle}

\begin{document}
\title{Cross-coupling effects in circuit-QED stimulated Raman adiabatic passage}

\author{A. Veps\"al\"ainen and G. S. Paraoanu}

\address{Department of Applied Physics, Aalto University, Puumiehenkuja 2 B, 02150, Espoo, Finland}

\ead{antti.vepsalainen@aalto.fi}

\begin{abstract}
Stimulated Raman adiabatic passage is a quantum protocol that can be used for robust state preparation in a three-level system. It has been commonly employed in quantum optics, but recently this technique has drawn attention also in circuit quantum electrodynamics. The protocol relies on two slowly varying drive pulses that couple the initial and the target state via an intermediate state, which remains unpopulated. Here we study the detrimental effect of the parasitic couplings of the drives into transitions other than those required by the protocol. The effect is most prominent in systems with almost harmonic energy level structure, such as the transmon. We show that under these conditions in the presence of decoherence there exists an optimal STIRAP amplitude for population transfer.
\end{abstract}

\section{Introduction}
Adiabatic pulse sequences can be used to transfer population between the states of a quantum system. As opposed to direct Rabi pulses, adiabatic protocols are not sensitive to the shape or frequency of the pulses and can therefore be used to realize robust quantum state preparation \cite{vitanov_review,stirap_sorin}. The downside of the adiabatic methods is their inherent slowness: population transfer is based on the adiabatic theorem \cite{adiabatic_theorem}, according to which the system remains in its instantenous eigenstate as long as the Hamiltonian governing the evolution of the system changes slow enough.

Historically, adiabatic methods have been used in quantum optics \cite{stirapfirst}, but recently they have started to gain traction also in other fields requiring accurate control of quantum state, such as circuit quantum electrodynamics (QED) where stimulated Raman adiabatic passage (STIRAP) has been studied theoretically \cite{stirap_photonics} and experimentally \cite{stirap_ours}. In the experiments it has been shown that STIRAP can be used to prepare the system to the second excited state with high fidelity while being robust to the control parameters. However, when the process fidelity is further optimized an additional problem arises. Unlike atomic systems, in circuit QED the energy level structure of the devices is often relatively close to a harmonic oscillator. This is especially true for a transmon \cite{transmon_PRA2007}, which is a prominent candidate for being a building block of a quantum computer. Consequently, the drive pulses are coupled off-resonantly to all other transitions. If the drive pulses are strong, the parasitic couplings start to play a detrimental role in the efficiency of the population transfer. This factor limits the amplitude of all the driving pulses, but it is especially important in the context of STIRAP protocol.

\subsection{STIRAP}
In a three-level system STIRAP can be used to transfer population from the ground state $\ket{0}$ to the second excited state $\ket{2}$ without having excitations in state $\ket{1}$ during the process. In this situation it is possible to realize 0 \--- 2 population transfer even in systems where that would otherwise be a forbidden transition, as is the case of the  transmon. The STIRAP protocol consists of two microwave drives that are coupled into the  0 \--- 1 and 1 \--- 2 transitions. In the adiabatic framework, the drives modify the system Hamiltonian such that its instantenous eigenstates slowly change. By properly designing the drives, it is possible to ensure that the initial state of the system evolves adiabatically to the desired target state. If the adiabatic condition of slow evolution is satisfied, the system does not get excited away from this state, and as a result it reaches the target state at the end of the protocol. We can see this by looking at the Hamiltonian of a three-level ladder system in a doubly-rotating frame, rotating with the two drives

\begin{equation}
\label{eq:three_level_hamiltonian}
H = \frac{\hbar}{2}\begin{bmatrix}
0 & \Omega_{0,1}(t) & 0 \\
\Omega_{0,1}(t) & 2\delta_{0,1} & \Omega_{1,2}(t) \\
0 & \Omega_{1,2}(t) & 2(\delta_{0,1}+\delta_{1,2})
\end{bmatrix}.
\end{equation}
The amplitudes of the drives are given by $\Omega_{0,1}(t)$ and $\Omega_{1,2}(t)$ and they are detuned from their corresponding transitions by $\delta_{i,j} = \omega_{i,j}^{\rm (d)} - \omega_{i,j}$, where $\omega_{i,j}^{\rm (d)}/(2\pi)$ are the drive frequencies and $\omega_{i,j}/(2\pi)$ are the system transition frequencies. The Hamiltonian can be diagonalized yielding the instantaneous eigenstates of the system

\begin{equation}
\begin{aligned}
\ket{+} =& \sin\Phi\ket{B} + \cos\Phi\ket{1}, \\ 
\ket{-} =& \cos\Phi\ket{B} - \sin\Phi\ket{1}, \\
\ket{D} =& \cos\Theta\ket{0} - \sin\Theta\ket{2},
\end{aligned}
\label{eq:stirap_eig}
\end{equation}
with eigenvalues
\begin{equation}
\begin{aligned}
\hbar\omega_+ =& \hbar \left(\delta_{0,1} + \sqrt{\delta_{0,1}^2 + \Omega_{0,1}(t)^2 + \Omega_{1,2}(t)^2}\right)/2, \\
\hbar\omega_- =& \hbar\left(\delta_{0,1} - \sqrt{\delta_{0,1}^2 + \Omega_{0,1}(t)^2 + \Omega_{1,2}(t)^2}\right)/2, \\
\hbar\omega_D =& 0.
\end{aligned}
\end{equation}
The bright state is $\ket{B} = \sin\Theta\ket{0} + \cos\Theta\ket{2}$ with the mixing angle defined as 
\begin{equation}
\label{eq:mixing_angle}
\tan\Theta = \Omega_{0,1}(t)/\Omega_{1,2}(t)
\end{equation}
while
\begin{equation}
\tan\Phi = \frac{\sqrt{\Omega_{0,1}(t)^2 + \Omega_{1,2}(t)^2}}{\sqrt{\Omega_{0,1}(t)^2 + \Omega_{1,2}(t)^2 + \delta_{0,1}^2} + \delta_{0,1}}.
\end{equation}
Here we have assumed that $\delta_{0,1} + \delta_{1,2} = 0$, realizing the two-photon resonance condition. Now we can see that as the ratio $\Omega_{0,1}(t)/\Omega_{1,2}(t)$ changes from $0$ to $\infty$, the mixing angle $\Theta$ goes from $0$ to $\pi/2$. As a result, the state $\ket{\rm D}$ changes from $\ket{0}$ to $\ket{2}$, realizing the required population inversion.

There are many possible choices for the shapes of the pulses $\Omega_{0,1}(t)$ and $\Omega_{1,2}(t)$ \cite{optimal_stirap_pulses}, but a common one is to take them as Gaussians

\begin{equation}
\begin{aligned}
\Omega_{0,1}(t) &= \Omega_{0,1}\exp{\left[-\frac{t^2}{2\sigma^2}\right]}, \\
\Omega_{1,2}(t) &= \Omega_{1,2}\exp{\left[-\frac{(t-t_{\rm s})^2}{2\sigma^2}\right]}.
\end{aligned}
\end{equation}
While $t_{\rm s}$ is negative the pulses are applied in an order where $\Omega_{1,2}$ pulse comes before $\Omega_{0,1}$ enabling the adiabatic population transfer. If the Hamiltonian \eqref{eq:three_level_hamiltonian} changes slowly enough, {\it i.e.},
\begin{equation}
\label{eq:ad_cond}
\sqrt{\Omega_{0,1}^2 + \Omega_{1,2}^2}\,\sigma \gg 1,
\end{equation}
the system does not get excited away from state $\ket{D}$. This means that for a fixed $\sigma$ it is favourable for the process to have as high drive amplitudes as possible in order to minimize the diabatic losses.

\section{Results}

\begin{figure}[tb]
\centering
\includegraphics[width=1.0\textwidth]{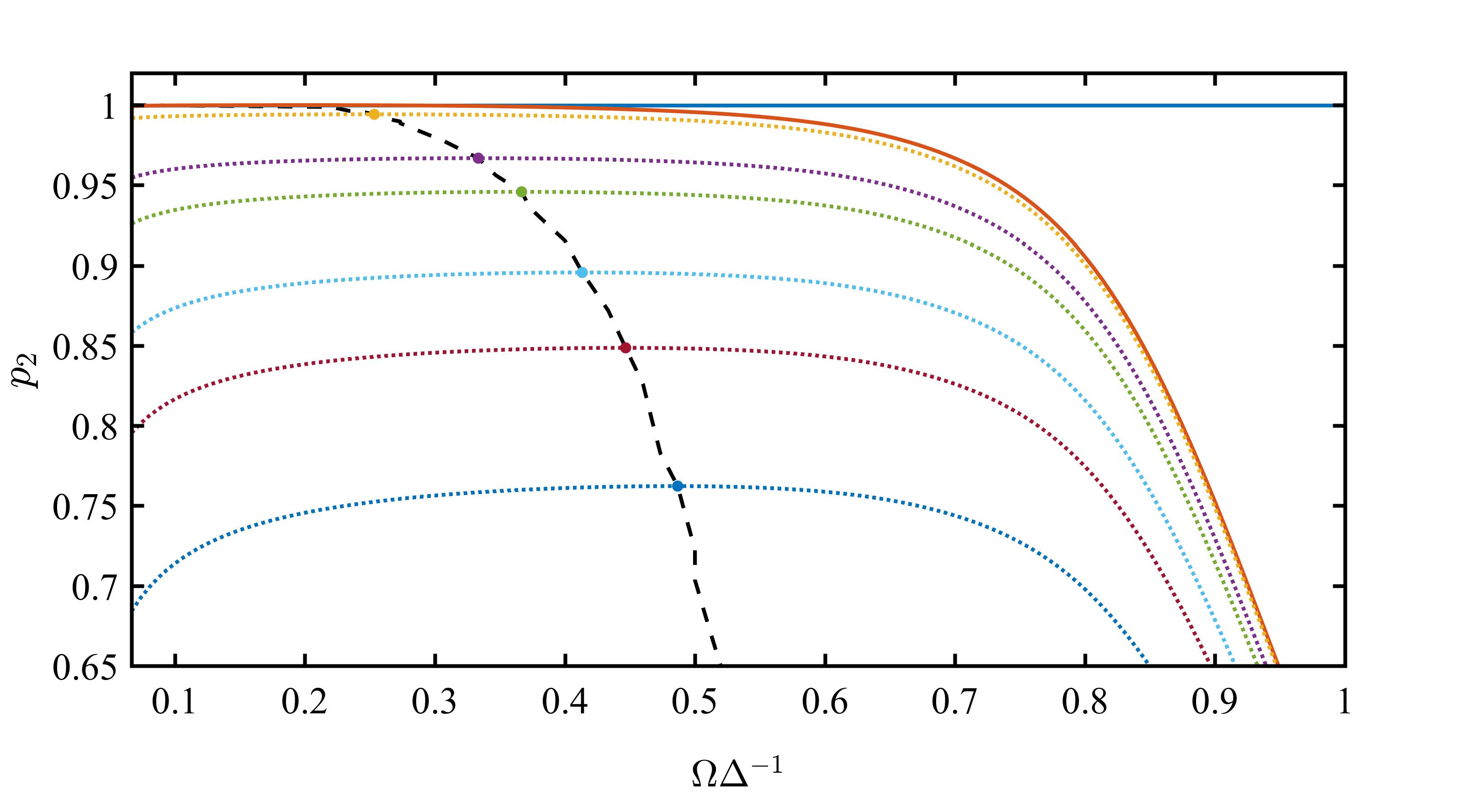}
\caption{Transferred population $p_2$ as a function of $\Omega\Delta^{-1}$, where $\Omega = \Omega_{0,1} = \Omega_{1,2}$, with cross-couplings (red solid line) and without cross-couplings (blue solid line). The dotted lines show the transferred populations in the presence of decoherence. The lines correspond to dimensionless decoherence rates $2\pi\Gamma\Delta^{-1}*10^{-4} = \{1, 10, 20, 30, 50, 100 \}$, starting from the highest line. The dashed black line shows the optimum $\Omega_{0,1}\Delta^{-1}$ for different values of $\Gamma$.}
\label{fig:A_vs_sigma}
\end{figure}

Until this point we have neglected the fact that both drive pulses couple off-resonantly also to the other transitions. In order to understand how this affects STIRAP, we start with the full Hamiltonian

\begin{equation}
\label{eq:full_hamiltonian}
H_{\rm full} = \hbar\sum_{k=1}^2\left( k\omega_{0,1} - (k-1)\Delta\right)\ket{k}\bra{k} + \frac{\hbar}{2}\sum_{k=0}^1\sum_{l=0}^1 \left(\Omega_{k,k+1}(t){\rm e}^{-i\omega_{k,k+1}^{\rm (d)}t}\ket{l+1}\bra{l} + {\rm H.c} \right),
\end{equation}
where $\Delta$ is the energy anharmonicity of the system. Note that this reduces to Eq. \eqref{eq:three_level_hamiltonian} under the rotating wave approximation given that $\Delta \gg \Omega_{k,k+1}$ for all $k$. If that is not the case, it is informative to consider the Hamiltonian describing only the parasitic couplings. Under the rotating wave approximation in the frame moving with the parasitic drives, we get

\begin{equation}
\label{eq:three_level_parasitic}
H_{\rm p} = \frac{\hbar}{2}\begin{bmatrix}
0 & \Omega_{1,2}(t) & 0 \\
\Omega_{1,2}(t) & -2\Delta + 2\delta_{1,2} & \Omega_{0,1}(t) \\
0 & \Omega_{0,1}(t) & 2(\delta_{0,1}+\delta_{1,2})
\end{bmatrix}.
\end{equation}
If the two-photon resonance condition $(\delta_{0,1} = -\delta_{1,2})$ is satisfied, the above Hamiltonian describes a STIRAP process from state $\ket{2}$ to state $\ket{0}$ with the roles of the drives reversed. As a result, the parasitic couplings create a competing STIRAP process, though with a significant single photon detuning. However, this process is not negligible, because STIRAP is robust against the single-photon detuning. Ironically, one of the main advantages of STIRAP here acts against it. Note that as long as $\Delta -\delta_{1,2} \gg \Omega_{1,2}(t)$ and $\Delta + \delta_{0,1} \gg \Omega_{0,1}(t)$ there are no diabatic excitations to state $\ket{1}$ even though the 1 \--- 2 drive is acting on the 0 \--- 1 transition and vise versa.

In order to understand the combined effect of the Hamiltonians \eqref{eq:three_level_hamiltonian} and \eqref{eq:three_level_parasitic} we numerically solve the time-evolution of the system under the full Hamiltonian \eqref{eq:full_hamiltonian} from
\begin{equation}
\label{eq:master_eq}
\dot{\rho}(t) = -\frac{i}{\hbar}[\rho(t),H_{\rm full}],
\end{equation}
with $\rho(-\infty) = \ket{0}\bra{0}$. For simplicity we assume that $\Omega_{0,1} = \Omega_{1,2} = \Omega$, $t_{\rm s} = -1.5\sigma$ and $\delta_{0,1} = \delta_{1,2} = 0$. With these conventions Eq. \eqref{eq:ad_cond} becomes $\Omega\sigma \gg 1$ and thus we can introduce a dimensionless parameter $a = \Omega\sigma$ which ideally determines the transfer fidelity of the STIRAP process. This is demonstrated in Fig. \ref{fig:A_vs_sigma}, where we show the population $p_2 = \bra{2}\rho(\infty)\ket{2}$ for $a = 3\pi$ when $\Omega$ and $\sigma$ are varied while keeping the anharmonicity $\Delta$ constant. The blue line describes the case without parasitic couplings. As expected, the fidelity of the process does not depend on the value of $\Omega$ because $\sigma$ is also changed accordingly. However, when the parasitic couplings are turned on (red line), the fidelity drops for larger values of $\Omega$ because the competing STIRAP process gets stronger as $\Omega$ approaches $\Delta$. Thus it would appear that it is optimal to make $\Omega$ as small as possible while allowing $\sigma$ to become longer. However, in the presence of decoherence the situation changes because longer $\sigma$ allows decoherence to affect more, suggesting that there exists an optimal value of $\sigma$ and $\Omega$ as a function of the decoherence rates. This is indeed the case and in Fig. \ref{fig:A_vs_sigma} we show $p_2$ for several different values of the energy relaxation rate $\Gamma$, where $p_2$ is solved from the Lindblad equation

\begin{equation}
\label{eq:lindblad}
\dot{\rho}(t) = -\frac{i}{\hbar}[\rho(t),H_{\rm full}] + \mathcal{L}[\rho],
\end{equation}
with  $\mathcal{L}[\rho] = -\Gamma \rho_{22} |2\rangle\langle 2| - (\Gamma \rho_{11} - \Gamma \rho_{22}) |1\rangle\langle 1| + \Gamma \rho_{11} |0\rangle\langle 0|$. The dashed black line shows the optimal value for the normalized $\Omega$. It is apparent that it is beneficial to use larger $\Omega$ for systems with higher decoherence rates. For example, in a typical transmon circuit we can take $\Delta/(2\pi) = 300$ MHz and $\Gamma = 0.5$ MHz, resulting in optimum of $\Omega/(2\pi) = 110$ MHz with $\sigma = 14$ ns.

\section{Conclusions}
We have shown that in systems with low anharmonicity the drives used in STIRAP protocol couple also to other transitions in a way that cannot be fully neglected. Even though in systems without decoherence the effect can be avoided by making the amplitudes of the drives very small and the duration of the pulses long, that is not optimal in general. We have demonstrated that the optimal value of the drive amplitude depends on the decoherence rate of the system, which we have numerically solved for several different relaxation rates. We believe that these results are useful when designing experiments exploiting STIRAP.

\section*{Acknowledgements}
We thank V\"ais\"al\"a Foundation, Academy of Finland CoE LTQ (project 250280), and
Centre for Quantum Engineering at Aalto University for financial support.

\section*{References}

\bibliographystyle{iopart-num}
\bibliography{ref}

\providecommand{\newblock}{}
\begin{thebibliography}{1}
\expandafter\ifx\csname url\endcsname\relax
  \def\url#1{{\tt #1}}\fi
\expandafter\ifx\csname urlprefix\endcsname\relax\def\urlprefix{URL }\fi
\providecommand{\eprint}[2][]{\url{#2}}

\bibitem{vitanov_review}
Vitanov N~V, Rangelov A~A, Shore B~W and Bergmann K 2017 {\em Rev. Mod.
  Phys.\/} {\bf 89}(1) 015006
  \urlprefix\url{https://link.aps.org/doi/10.1103/RevModPhys.89.015006}

\bibitem{stirap_sorin}
Falci G, Di~Stefano P~G, Ridolfo A, D'Arrigo A, Paraoanu G~S and Paladino E
  2017 {\em Fortschritte der Physik\/} {\bf 65} 1600077--n/a ISSN 1521-3978
  1600077 \urlprefix\url{http://dx.doi.org/10.1002/prop.201600077}

\bibitem{adiabatic_theorem}
Born M and Fock V 1928 {\em Zeitschrift f{\"u}r Physik\/} {\bf 51} 165--180
  ISSN 0044-3328 \urlprefix\url{http://dx.doi.org/10.1007/BF01343193}

\bibitem{stirapfirst}
Gaubatz U, Rudecki P, Schiemann S and Bergmann K 1990 {\em The Journal of
  Chemical Physics\/} {\bf 92} 5363--5376
  \urlprefix\url{http://scitation.aip.org/content/aip/journal/jcp/92/9/10.1063/1.458514}

\bibitem{stirap_photonics}
Veps{\"a}l{\"a}inen A, Danilin S, Paladino E, Falci G and Paraoanu G~S 2016
  {\em Photonics\/} {\bf 3} ISSN 2304-6732
  \urlprefix\url{http://www.mdpi.com/2304-6732/3/4/62}

\bibitem{stirap_ours}
Kumar K~S, Veps{\"a}l{\"a}inen A, Danilin S and Paraoanu G~S 2016 {\em Nature
  Communications\/} {\bf 7}
  \urlprefix\url{http://dx.doi.org/10.1038/ncomms10628}

\bibitem{transmon_PRA2007}
Koch J, Yu T~M, Gambetta J, Houck A~A, Schuster D~I, Majer J, Blais A, Devoret
  M~H, Girvin S~M and Schoelkopf R~J 2007 {\em Phys. Rev. A\/} {\bf 76} 042319
  \urlprefix\url{http://journals.aps.org/pra/abstract/10.1103/PhysRevA.76.042319}

\bibitem{optimal_stirap_pulses}
Vasilev G~S, Kuhn A and Vitanov N~V 2009 {\em Phys. Rev. A\/} {\bf 80}(1)
  013417 \urlprefix\url{http://link.aps.org/doi/10.1103/PhysRevA.80.013417}

\end{thebibliography}

\end{document}